\documentclass[10pt,twocolumn]{article}

\usepackage[utf8]{inputenc}
\usepackage[english]{babel}
\usepackage{amsmath}
\usepackage{amsfonts}
\usepackage{amssymb}
\usepackage{mathpazo} 
\usepackage[scaled]{helvet}
\usepackage[T1]{fontenc}
\usepackage{url}
\usepackage{verbatim}
\usepackage{float}
\usepackage{caption}
\usepackage{subcaption}
\usepackage{graphicx}
\usepackage{xcolor}
\usepackage{booktabs}
\usepackage{algorithm}
\usepackage[noend]{algpseudocode}
\usepackage{changepage}
\usepackage[normalem]{ulem}
\usepackage{soul}
\usepackage[left=48pt,%
                right=42pt,%
                top=42pt,%
                bottom=60pt,%
                headheight=15pt,%
                headsep=10pt,%
                letterpaper,twoside]{geometry}%
                
\usepackage[labelfont={bf,sf},%
                labelsep=period,%
                figurename=Fig.,%
                singlelinecheck=off,%
                justification=RaggedRight]{caption}
                
\usepackage[numbers,sort&compress]{natbib}
 \title {Controlling intense, ultrashort, laser-driven relativistic mega-ampere  electron fluxes by a modest, static magnetic field} 

\usepackage{authblk}
\author[1]{Anandam Choudhary}
\author[2]{Trishul Dhalia}
\author[1]{C. Aparajit}
\author[1]{Amit D. Lad}
\author[1]{Ankit Dulat}
\author[1]{Yash M. Ved}
\author[2]{Rohit Juneja}
\author[2,*]{Amita Das}
\author[1,*]{G.Ravindra Kumar}

\affil[1]{Tata Institute of Fundamental Research, 1 Homi Bhabha Road, Colaba, Mumbai 400 005, India}
\affil[2]{Indian Institute of Technology, Delhi }
\affil[*]{Corresponding author: amita@iitd.ac.in; and grk@tifr.res.in}

\date{}

\begin{document}

\twocolumn[
  \begin{@twocolumnfalse} 
    
    \maketitle
    
    \begin{abstract}

The guiding and control of ultrahigh flux, femtosecond relativistic electron pulses through solid density matter is of great importance for many areas of high energy density science. Efforts so far include the use of magnetic fields generated by the propagation of the electron pulse itself or the application of hundreds of Tesla magnitudes, pulsed external magnetic fields driven by either short pulse lasers or electrical pulses. Here we experimentally demonstrate the guiding of hundreds of keV mega-ampere electron pulses in a  magnetized neodymium solid that has a very modest, easily available static field of 0.1 tesla. The electron pulses driven by an ultrahigh intensity, 30 femtosecond laser are shown to propagate beam-like, a distance as large as 5 mm in a high Z target (neodymium), their collimation improved and flux density enhanced nearly by a factor of 3. Particle-in-cell simulations in the appropriate parameter regime match the experimental observations. In addition, the simulations predict the occurrence of a novel, near-monochromatic feature towards the high energy end of the electron energy spectrum, which is tunable by the applied magnetic field strength. These results may prove valuable for fast electron beam-driven radiation sources, fast ignition of laser fusion, and laboratory astrophysics.      
    
    \vspace{2mm}
    \end{abstract}
    
  \end{@twocolumnfalse}
]
\begin{figure*}[!ht]
   \centering
    \includegraphics[width=0.9\linewidth]{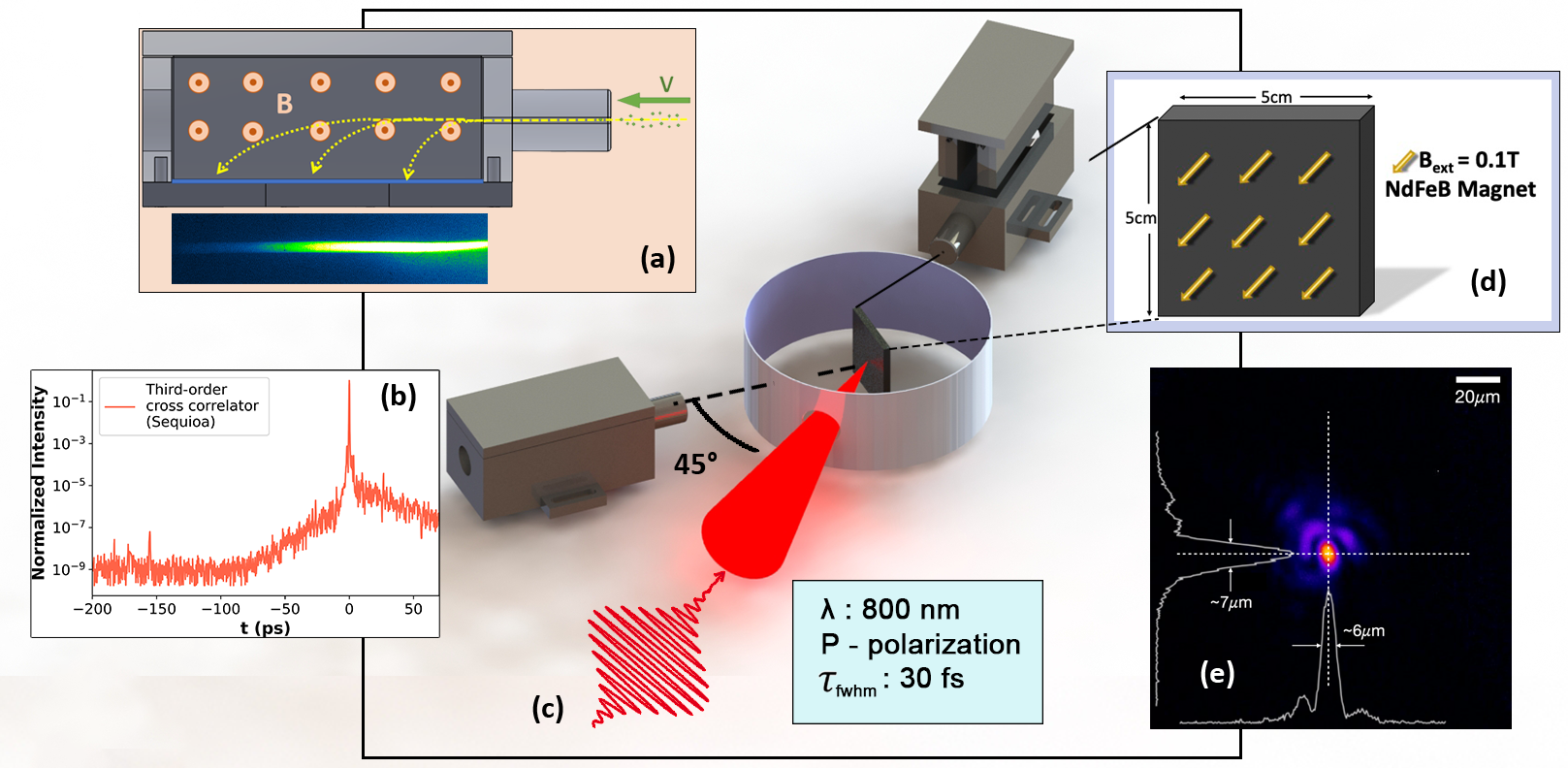}
   \caption{(a) Trajectory of electrons inside ESM(electron spectrometer) magnetic field. Electrons trace on the image plate is also shown. (b) contrast measurement of the laser pulse (pulse width - 30fs). (c) Illustration of the ESM and angular distribution measurement setup. (d) Neodymium magnetic target with 0.1 Tesla magnetic field directed normal to the surface. (e) The laser pulse focal spot (6 $\mu m$ FWHM ) measured during the experiment.}
   \label{fig:Exp_Setup}
\end{figure*}
 \begin{figure}[!ht]
   \centering
    \includegraphics[width=1.0\linewidth]{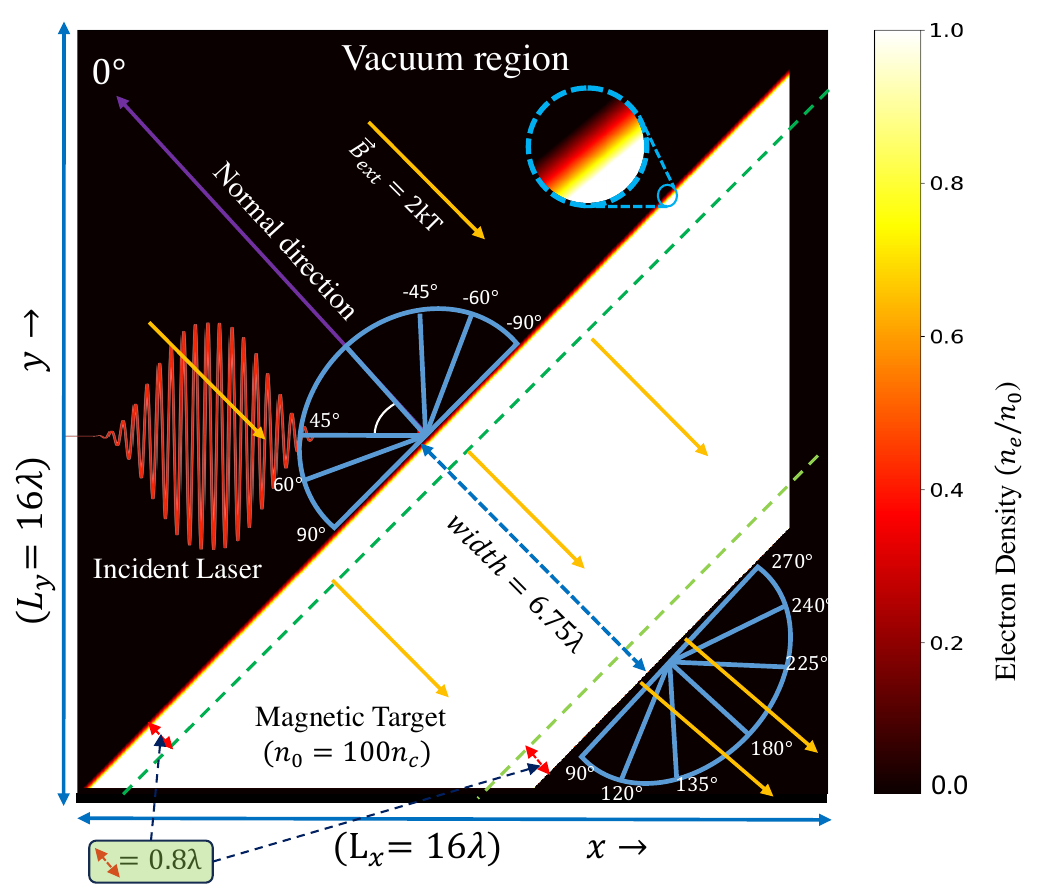}
   \caption{Simulation setup for electron angular fluxes measurement on the OSIRIS 4.0 platform. The magnetic target was designed by applying an external magnetic field (2kT) normal to the surface. Intense laser of $\lambda$= 800nm was incident obliquely at 45 degrees from the normal direction. The density scale length in the front surface was chosen $0.2\lambda$. For angular distribution in front and rear direction, electrons having energy $100$ keV or more within $0.8\lambda$ from front and back surfaces (dotted green lines) have been collected.}
   \label{fig:Setup}
\end{figure}
\begin{figure*}[!ht]
    \begin{subfigure}{.47\linewidth}
        \centering
		\includegraphics[width=\columnwidth]{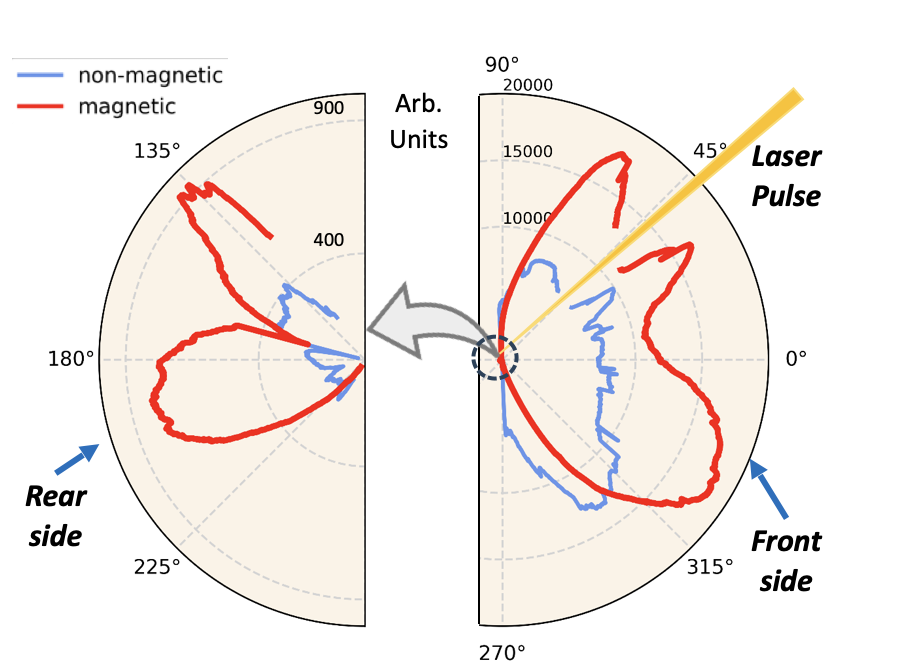}
		\caption{}
		\label{fig:Ang_Exp}
    \end{subfigure}
    \begin{subfigure}{.58\linewidth}
        \centering
        \includegraphics[width =\columnwidth]{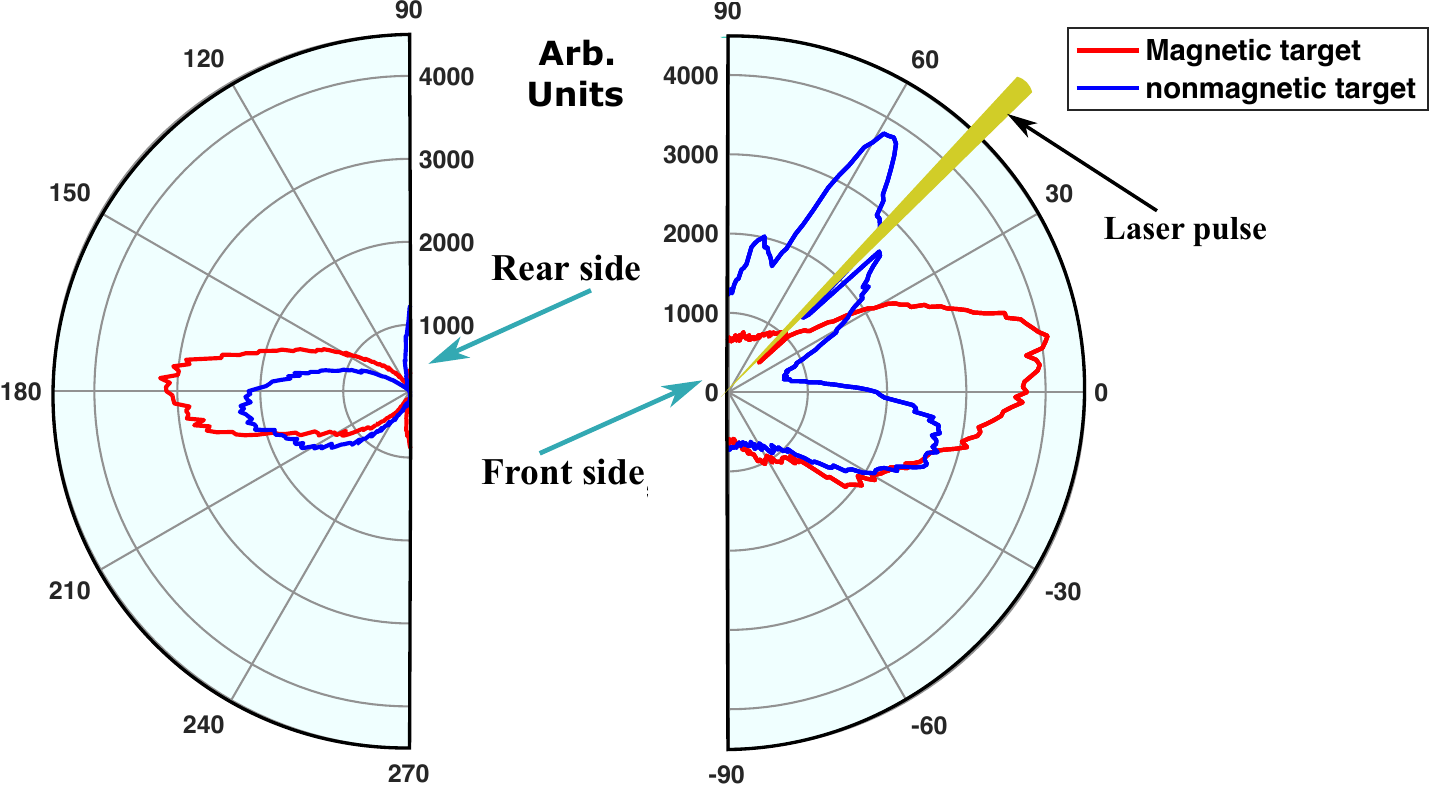}
        \caption{}
        \label{fig:Ang_Sim}
    \end{subfigure}
    \caption{(a) Electron angular distributions for magnetic and non-magnetic targets from experiments (left: rear side and right: front side) (b) Electron angular distributions from simulations} 
    
\end{figure*}

\begin{figure*}[!ht]
   \begin{subfigure}{.47\linewidth}
        \centering
		\includegraphics[width=\columnwidth]{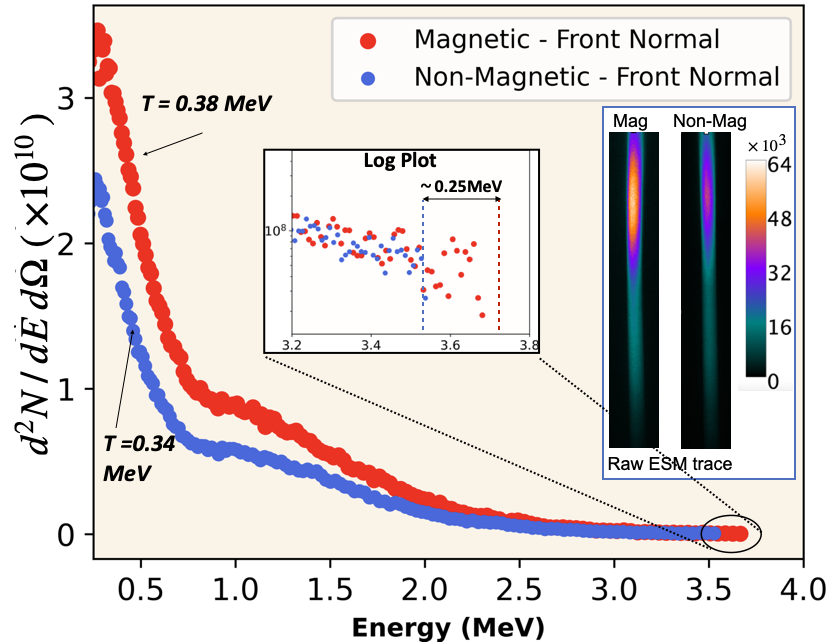}
		\caption{}
		\label{fig:Spectrum}
    \end{subfigure}\hspace{2mm}
    \begin{subfigure}{.5\linewidth}
        \centering
        \includegraphics[width = \columnwidth]{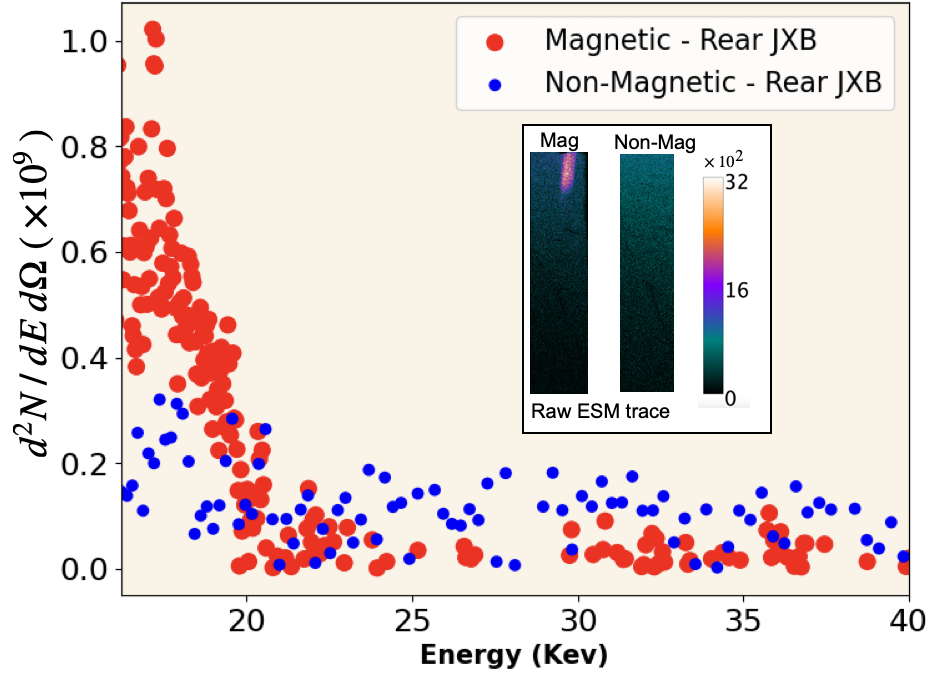}
        \caption{}
        \label{fig:Rear_spectrum}
    \end{subfigure}
    \caption{(a) Electron energy spectrum at the front normal side for the magnetic and non-magnetic target. The inset (log scale) shows 0.25 MeV more energy cut-off for the case of the magnetic target. ESM traces are also shown  (b) Electron energy spectrum at the rear side. ESM traces of electrons on image plates along the rear laser direction ($\vec{J}\times \vec{B}$). At the rear side, the non-magnetic target shows no clear signal(mostly noise and background). The magnetic target shows a clear signal that is the distinct signature of energetic electrons passing through the 5mm thick Neodymium magnet.} 
    
\end{figure*}

\begin{figure*}[!ht]
   \centering
    \includegraphics[width=1.0\linewidth]{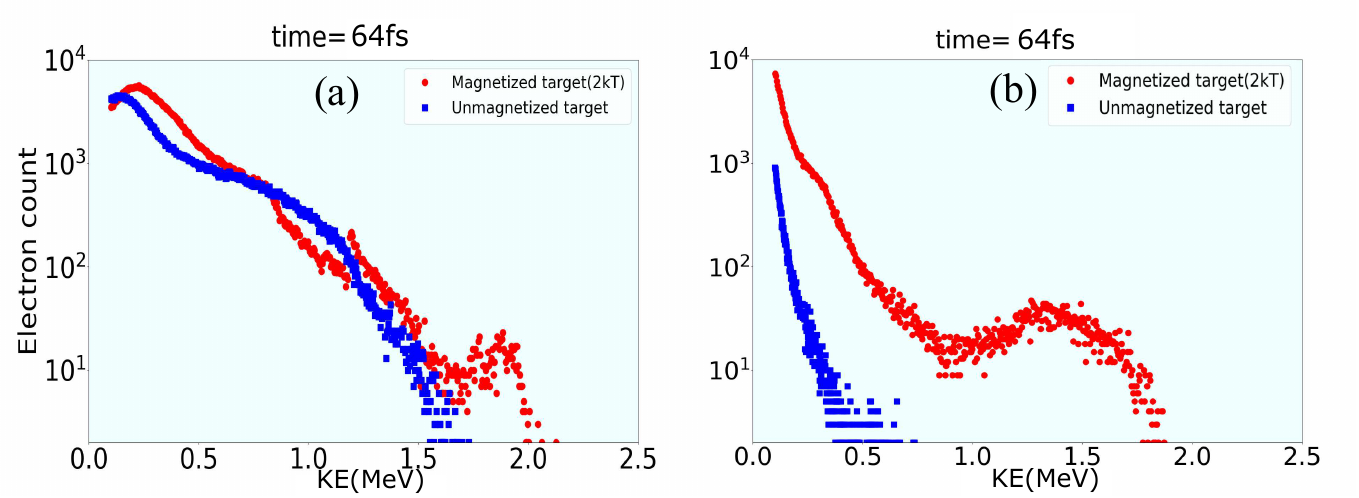}
   \caption{Energy distribution of electrons obtained from simulations for (a) front surface (b) rear surface after the laser has left the simulation box at t=64 fs }
   \label{fig:Sim_Spectrum}
\end{figure*}

\begin{figure}[!ht]
    \centering
    \includegraphics[width=1.0\linewidth]{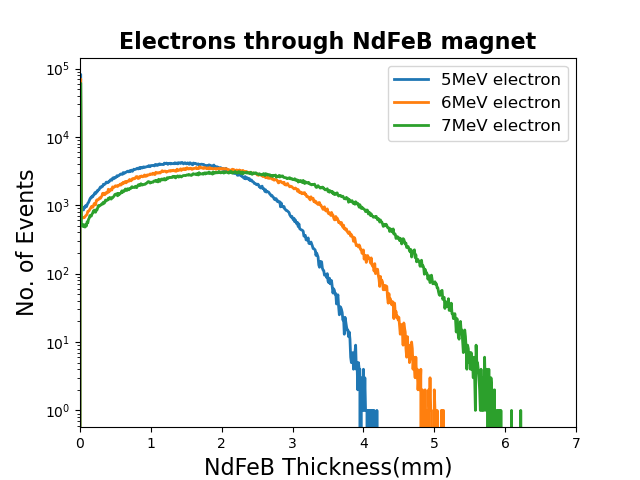}
    \caption{GEANT4 simulation results show events of electron propagation events through the neodymium magnet. It can be seen that only electrons with energy at least 6 MeV can only pass through the 5mm thick magnet}
    \label{fig:geant4}
\end{figure}

\begin{figure*}[!ht]
   \centering
    \includegraphics[width=0.95\linewidth]{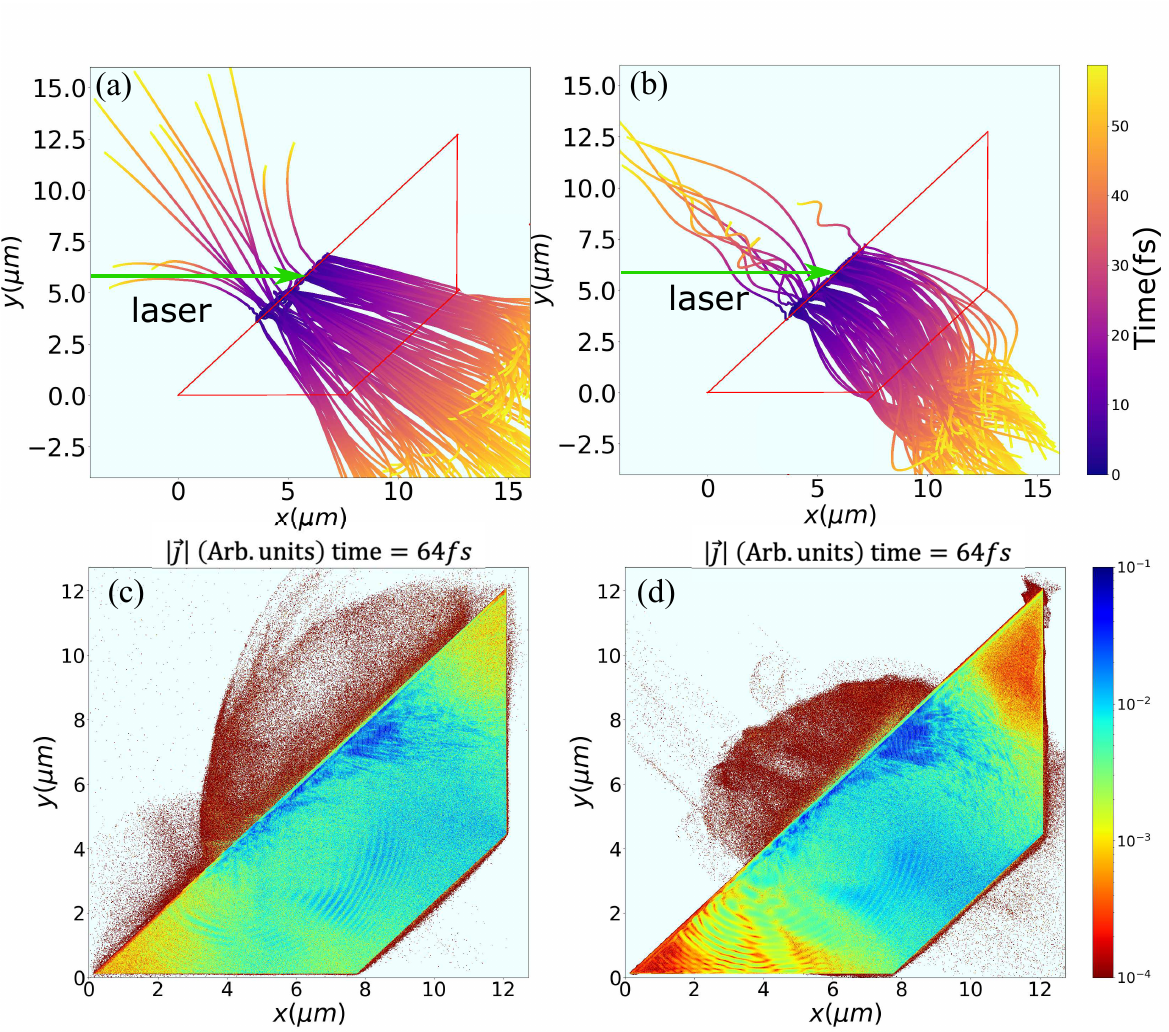}
   \caption{Particle trajectory of the randomly chosen 2000 particles from the front surface of the target has been shown up to 64 fs for (a) non-magnetized and (b) magnetized target. Net current density generated at 64 fs has been plotted when the laser has left the simulation box for (c) non-magnetized (d) magnetized target}
   \label{fig:Current_trajectory_sim}
\end{figure*}

\begin{figure*}[!ht]
    \centering
    \includegraphics[width=0.95\linewidth]{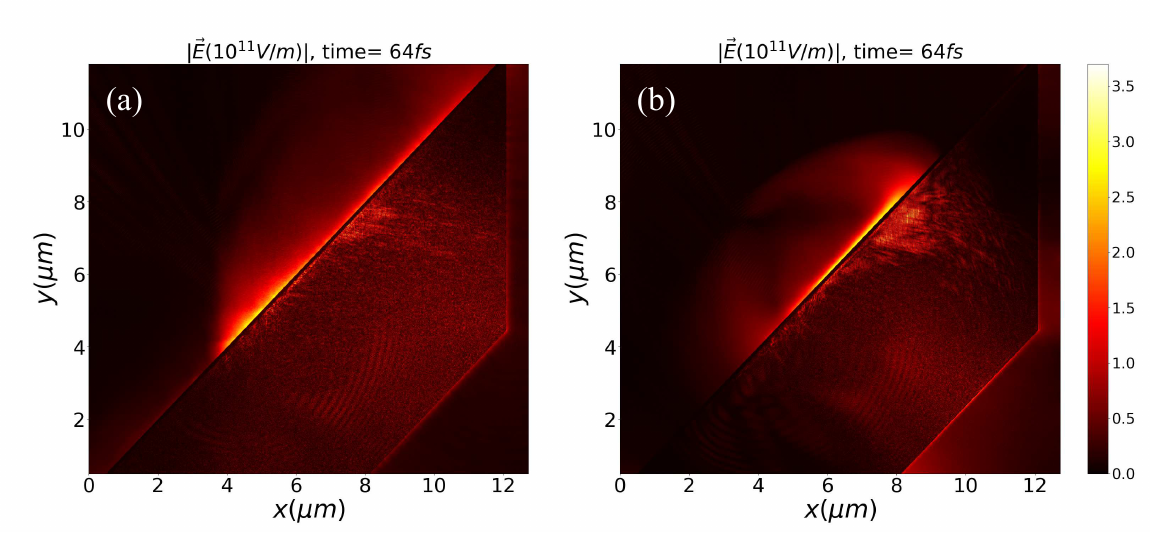}
    \caption{Figure illustrates the generated sheath electric  field at the surface of (a) non-magnetic target (b) magnetic target after the laser has passed the simulation box after $64 fs$.}
    \label{fig:sheath_field}
\end{figure*}

\begin{figure}[!ht]
    \centering
    \includegraphics[width=0.95\linewidth]{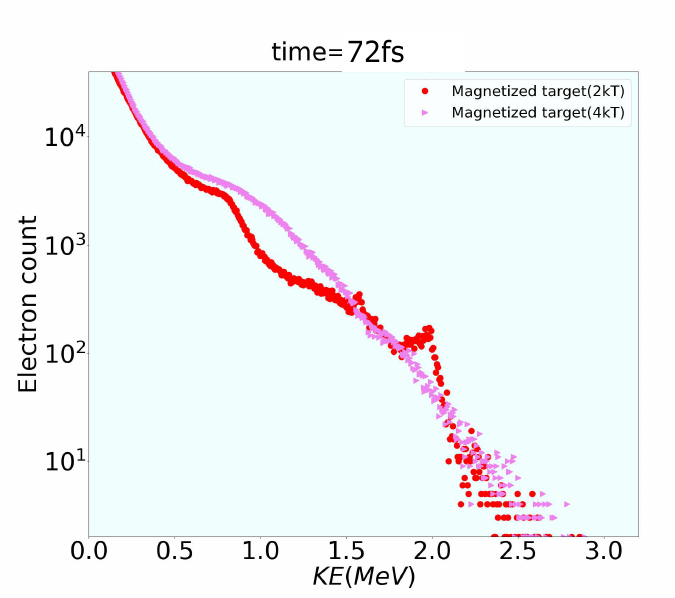}
    \caption{Figure depicts energy distribution of electrons at different times for external magnetic field of 2 kT and 4 kT.}
    \label{fig:cut_off_energy}
\end{figure}

\begin{figure*}
    \centering
    \includegraphics[width=0.95\linewidth]{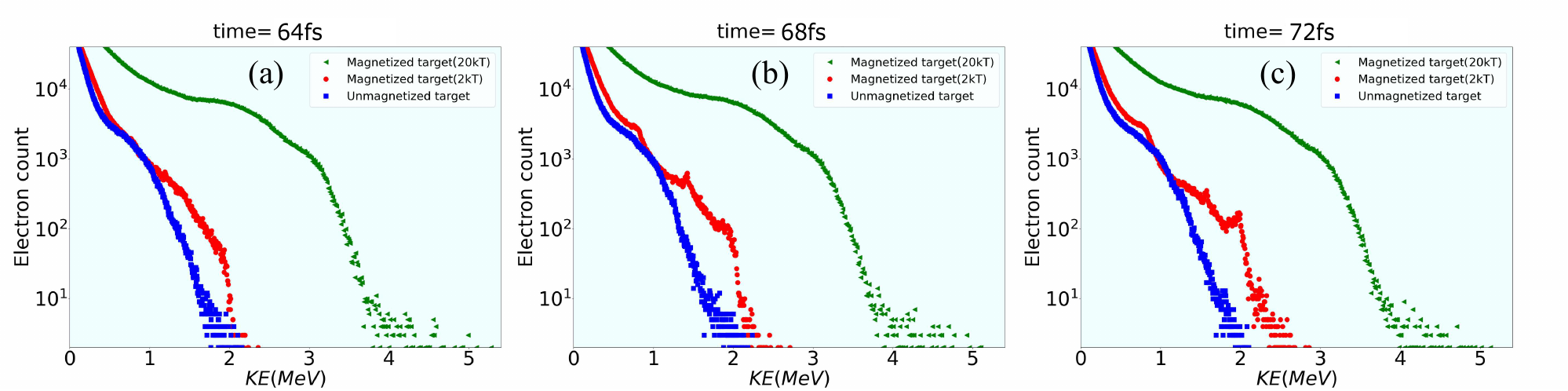}
    \caption{(a,b,c) demonstrate the energy distribution of electrons in the whole simulation box for three different targets (non-magnetic, magnetic $(2kT)$ and magnetic $(20kT)$ at three different times, respectively. }     
    \label{fig:energy_dist_time_evol}
\end{figure*}

\begin{figure}
    \centering
    \includegraphics[width=0.95\linewidth]{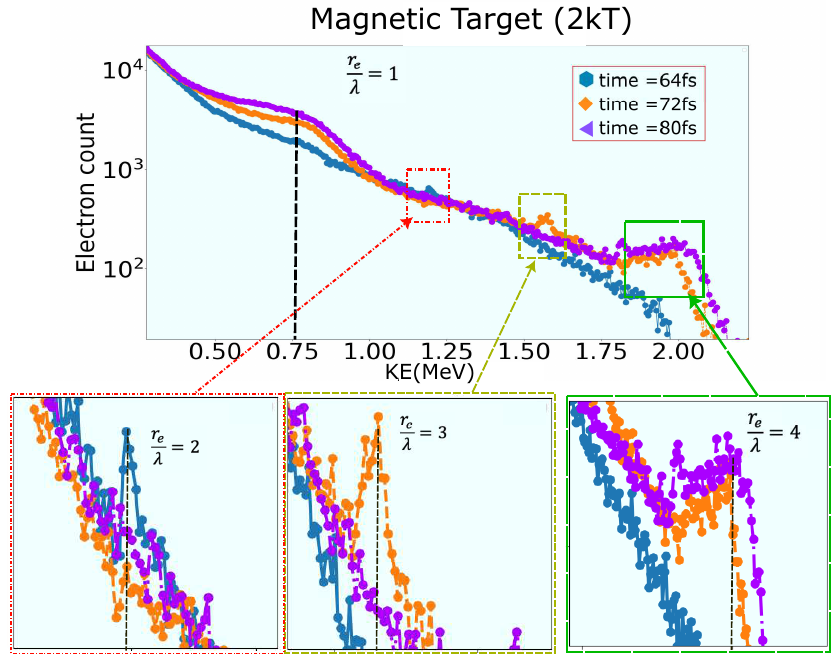}
    \caption{Peaks in the energy spectrum at the locations of integer values of relativistic modified $r_e/\lambda$ ratio.}     
    \label{fig:mono_energy}
\end{figure}

\section{Introduction}
 High-intensity, short, and ultrashort pulses can drive matter to extreme states in a laboratory environment \cite{Kaw2017, drake_2006}. This is a hotly investigated area with a huge basic science impact as well as myriad exciting applications in laboratory astrophysics, electron and ion acceleration and bright hard X-ray sources which in turn can advance condensed matter science, particle and nuclear physics, chemistry, and biology \cite{drake_2006, Bulanov_RevModPhys.94.045001}.  The crucial players in this physics are the relativistic, megaampere pulses of electrons generated by the driving visible/infrared laser pulses. It is natural therefore that much attention is focused on the mechanisms to generate, tailor, and control the generation of these electron pulses and their transport through the surrounding plasma environment. 
 
 These large fluxes of relativistic electrons are typically generated in solid targets. Several innovative methods have increased the fluxes of these electrons namely the use of nanostructured surfaces \cite{rajeev_PhysRevLett.90.115002, Purvis2013}, sub-lambda gratings \cite{kahaly_PhysRevLett.101.145001, Lad2022, Macchi_10.1063/1.5013321, Fedeli_PhysRevLett.116.015001} etc. The high background density that the generated electron beam experiences, however, poses several complexities. The beam divergence,  shown to increase with laser intensity \cite{Green_PhysRevLett.100.015003} is an important parameter that needs to be controlled for the effective transfer of energy to secondary emissions. The beam is also prone to several instabilities in the dense surrounding plasma \cite{Kaw2017, Weibel_PhysRevLett.2.83, Pegoraro_1996} and this can seriously damage its integrity via filamentation in as short a transport length as a few tens of microns, shown in several computer simulations. In the initial stages of transport, the particles may be treated as ballistic but with loss of energy and increasing divergence in the transport, collisional as well as collisionless \cite{yabuuchi2009evidence} stopping can play a significant role. In applications like fast ignition of laser fusion \cite{Tabak_10.1063/1.870664} petawatt peak power, femtosecond/picosecond driven megaampere beams are required to deposit their energy in a very small (tens of micrometers) region, propagating through an imploded super-solid density target \cite{Vauzour_PhysRevLett.109.255002}.   This is an issue that continues to challenge us.
 
 How can we guide these mega fluxes of electrons through a solid?  One approach has been to use tailored targets.  For example, carbon nanotubes \cite{Chatterjee_PhysRevLett.108.235005} and metal nanochannels \cite{PKSingh_PhysRevSTAB.16.063401} have been employed to enhance the generation as well as transport of large currents over lengths much larger than the typical filamentation lengths of a few microns. In fact, the nanotube target transports large currents over a distance as large as a millimeter, the structure later theorized to prevent the onset of the deleterious Weibel instability \cite{Mishra}. An interesting innovation has been the use of the resistivity gradient in the medium surrounding the plasma channel, called resistive collimation \cite{Bell_PhysRevLett.91.035003}, to guide the electrons \cite{Kar_PhysRevLett.102.055001, Ramakrishna_PhysRevLett.105.135001}, going so far to maximize energy deposition in a specified ignition volume \cite{Robinson_PhysRevLett.108.125004}. Tailoring the density gradient has also been suggested as a possible way to guide electrons \cite{yadav2009anomalous}. While all these are internal to the target, large external magnetic fields (10$^3$ tesla) have been postulated in simulations \cite{Cai_10.1063/1.4812631, Bailly-Grandvaux2018} and even experimentally applied to imploded fusion cores \cite{Sakata2018} to reduce divergence and enhance coupling of the laser to the electrons. The availability of increasingly high magnetic fields in the laboratory in fact, leads to an interesting and altogether new magnetized regime of laser plasma interactions for which theoretical and simulation studies are being carried out  \cite{das2020laser, PhysRevE.105.055209,vashistha2020new,vashistha2021excitation,goswami2021ponderomotive}.

 In a  broader sense, the transport of high-velocity, large flux charged particles through the universe is subject to magnetic field environments in astrophysical objects as well as interstellar space. The results from laboratory studies may throw light on those aspects too.
 
For wider applicability, however, it is necessary to 
 significantly simplify the experimental requirements (both in the structure of the target and the external magnetic fields). Besides, the above-mentioned structures, designer resistivity gradients, and large external fields applied, while interesting in their own way, may not reveal the physics of the core transport process itself. In this paper, therefore we present the simple situation of a {\it modest},  {\it static} magnetic field and a  normal homogeneous solid to illustrate the rich possibilities that exist for the generation control and transport of large flux MeV electron pulses. We will provide a comparative discussion later but for now, it is enough to state that we use readily available, inexpensive permanent magnet neodymium metal targets (5 mm thick)  with an axial field of 0.1 Tesla and irradiate them with 30 fs, 800 nm, 1 $\times$10$^{19}$ W cm$^{-2}$  laser pulses.  We measure the fast electron energies, fluxes, and angular distributions at both the front and back of this 5  mm long target. We see clear signatures of collimation by the ambient magnetic field even though it is insignificant in magnitude to the self-generated field in the plasma, which is ~100s Tesla. We see enhanced fluxes of electrons at the target front, where they are created, and the emergence of a large flux even after 5 mm long travel through a high Z material like neodymium. Further, We observe clear steering of these beams by the sheath fields at the target-vacuum interfaces both at the front and the back. 
 
To support and explain our observations,  we carried out particle-In-Cell simulations using  the OSIRIS4.0 \cite{hemker2000particle,fonseca2002osiris,fonseca2008one} platform. These were carried out choosing the applied strength of the magnetic field considerably higher than the experimental value. This had to be done so that the collimation of charged particles could be observed at a comparatively shorter distance (constrained by the choice of simulation box size) than the experimental distance of 15 cm. This simply makes the gyro radius of the species tighter and any beam collimation noticeable within a short distance. However, appropriate care was taken to have the magnetic field strength low enough not to change the dispersion characteristics of the laser wave. The simulation compares well with the experimental observations discussed later in detail. Simulations also reveal a striking new feature, namely a quasi-monochromatic electron bunch whose peak energy as well as energy width are tunable by the applied magnetic field. We explain the physics by looking at the essential scales of electron transport and offer insights and prospects in relativistic electron transport through magnetized dense media. We would like to underline here that the magnetic field chosen in this experiment is very weak. The laser wave propagation inside the plasma continues to be governed by the un-magnetized dispersion relation as the gyro-frequency of  the electron species is much smaller than the laser frequency. The signature of this weak magnetic field in experiments is felt by the electrons that are observed much later when their trajectories have undergone evolution over several gyro periods.  
The other regime of strong magnetic fields, (for which the gyrofrequency of one or both the plasma species can be higher than the laser frequency) has attracted considerable interest lately (with the kilo Tesla level magnetic field generation in the laboratory, the gyrofrequency of electrons can be higher than the laser frequency for short pulse $CO_2$ lasers.)  For these cases, the propagation of the laser/EM wave itself in plasma gets modified. It follows the magnetized dispersion relation, which permits various pass bands even for an overdense plasma. Simulation and analytical studies in this domain have revealed interesting possibilities for laser beam transportation \cite{maity2021harmonic,dhalia2023harmonic,Juneja_2023,goswami2022observations}.

\section{Experiment}
The experiment was conducted with the Tata Institute of Fundamental Research 150 TW laser system. 5mm thick neodymium targets (magnetized and un-magnetized) were irradiated by 800 nm, 30 femtosecond p-polarised laser pulses focused to a $6\mu m$ spot (FWHM) by an $f/3$ off-axis parabolic mirror at $45^{\circ}$ incidence angle. Peak intensity of $~5\times$ 10$^{19}$ W/cm$^{2}$ was achieved with $10^{-8}$ intensity contrast at 20 picoseconds. The strength of the magnetic field in the magnetized target was 0.1 tesla directed normal to the surface.

To diagnose the distribution of electron fluxes, image plates(IP) were placed in a cylindrical geometry surrounding the target at the center (Fig.\ref{fig:Exp_Setup}). 10 layers of $11\mu m$ thick aluminum foil are used to cover image plates to block low energy electrons ($100 keV$) and other visible emissions. Image plates work on the principle of photo-stimulated luminescence \cite{Tanaka_10.1063/1.1824371} and record the electron flux information which can be read using image scanners.

We used electron spectrometers (ESM) (Fig.\ref{fig:Exp_Setup}) to measure electron energies and temperatures placed in two different directions (front normal and along the laser propagation at target rear which is $\vec{J}\times \vec{B}$). Each electron spectrum was obtained by averaging 15 laser shots. The experiment was performed in a vacuum chamber at $10^{-5}$ Torr pressure.


\section{Simulation}

 2D3V PIC simulations were performed using the OSIRIS4.0 framework to simulate the pertinent physics domain of the experimental findings. The schematic of the simulation study has been provided in Fig.2.  A p-polarised laser pulse of wavelength $\lambda =800 nm $ was incident obliquely at an angle of $45^\circ$  on the target plasma surface from the left boundary. The plasma consists of electrons and neutralizing heavy ions. The ions are assumed to remain static throughout the simulation.  The simulation box size was chosen as ($16 \lambda \times 16 \lambda$). The laser pulse duration consists of 5 laser cycles and the transverse spot size was about  $4.5\lambda$ ($3.6 \mu m$). A normalized amplitude $a_\circ = 6$ has been used in agreement with the experimental laser intensity of  $8\times 10^{19} W/cm^2 $. A fully ionized plasma target of density $100n_c$ and of longitudinal width $6.7\lambda$ ($7.7\mu m$) is considered. An exponentially sharp plasma density gradient profile of the form  $n=100n_c(\exp
((x-y)\ln(2)/L)-1)$ has been introduced at the front surface of the target for a density scale length of $L=0.2 \lambda$ ($160 nm$). 
 
 The simulations cannot be performed with the same choice of experimental parameters. The target dimension (5cm$\times$5cm$\times$5mm) with external magnetic field region of influence ($\sim10mm$ normal to the surface and $\sim50mm$ in transverse direction) is too large to simulate. For this reason, one needs to appropriately scale the experimental parameters. The attempt here is to simulate a physical regime that is similar to the experimental scenario.
 
 For the choice of 0.1 T applied external magnetic field, the electron gyrofrequency $ \omega_{ce} = 7.46\times 10^{-6}\omega_l$ is much smaller than the laser frequency $\omega_l$. The laser EM field thus experiences only the un-magnetized plasma response. The electrons that get influenced by the 0.1-tesla magnetic field of the target up to $\sim10mm$ from the target rear surface would, however, undergo several gyro-oscillations before finally hitting the detector. We measured the magnetic field strength and found it to be significant up to this distance. The time spent under the influence of the external field by even relativistic electrons moving in a helical path (gyro radius  $\sim10mm$) from the plasma target is about $\sim10^{-10}sec$ ($> 10 \omega_c^{-1}$) which is longer than the gyro-period. Thus, even this weak applied magnetic field has a significant role to play in defining the electron trajectories toward the detector. The choice of magnetic field in the simulations has been made keeping these two considerations in view.

  In our simulations, a magnetic field of 2kT has been chosen for most of the runs. Where except for some cases, 4kT and 20kT have also been chosen. While for 2kT and 4kT the laser obeys the un-magnetized dispersion relation, the magnetized dispersion relation for the laser is valid for 20 kT. The energetics and directional guidance of electrons have been measured at a distance of up to $5 \mu m$.
 For this value, the gyro-radius of the electrons is of the order of $\sim$ $\mu$m $\sim \lambda_l$, the laser wavelength.  For a kilo Tesla magnetic field, the laser continues to obey the unmagnetized dispersion relation as $\omega_{ce} < \omega_l$. The physics regime of laser-plasma interaction as considered in experiments has therefore not been altered. The direction of the external magnetic field has been chosen to be normal to the front surface of the target, to replicate the experimental situation. 
 
 For some of our simulations, we have chosen even higher values of magnetic fields (for which the laser propagation characteristics have to be governed by magnetized dispersion relation) to underline certain novel features that cannot be captured in this set of experiments. These observations have been made at a choice of the external magnetic field of 20kT corresponding to $\omega_{ce}=1.5\omega_l$.
 The results and observations of the simulations have been presented in the next section. The comparison with experimental findings shows similarities as well as differences which are expected due to the high value of the magnetic field chosen in the simulation. 
\section{Results and Discussion}

 Figures \ref{fig:Ang_Exp} and \ref{fig:Ang_Sim} depict the angular distribution of electron fluxes observed in the experimental and simulation data respectively, from both the front and rear surfaces of the target. Only fast electrons with energies exceeding $100 keV$ were detected at the IP for both magnetized and non-magnetized targets. The figures clearly demonstrate that the magnetized target exhibits a higher electron flux compared to the non-magnetized target in both experiments and simulations. The angular distribution of electrons, however, continues to remain relatively broad at both the front and back surfaces of the target for experiment in comparison to simulation. In simulations, there is a strong collimation of the electrons along the magnetic field direction since a very strong magnetic field has been employed. 

 The Electron Spectrometer (ESM) results, shown in Figure \ref{fig:Spectrum}, also exhibit similar trends. The effective temperature of hot electrons is comparable for both magnetized and non-magnetized cases. However, the cutoff energy on the tails of the spectrum is higher ($\sim$3.7 MeV compared to 3.5 MeV) for the magnetized targets. Figure \ref{fig:Rear_spectrum} displays the energy spectrum for both targets at the rear. No distinct signal (noise and background only) is observed for the non-magnetic target, while the magnetic target exhibits a clear and enhanced signal above the noise level. This represents the unmistakable signature of energetic electrons effectively traversing the 5mm thick Neodymium magnet enduring energy loss due to collisions and other processes, in the presence of an external static magnetic field. According to a GEANT4 simulation (Figure \ref{fig:geant4}), only electrons with energy as large as 6-7 MeV can only pass through this target, indicating that the electrons possess substantially higher energies inside the target than those measured by our spectrometer.
 
 The simulation Fig.\ref{fig:Sim_Spectrum} aligns with our results, demonstrating higher fluxes and energy cutoffs for the magnetized target, both at the front and rear. Fig.\ref{fig:Sim_Spectrum}(b) shows the gradual increase in electron counts with energy 1.2 MeV and more, measured at target rear. These electrons are transported from the front to the rear for the magnetic target, showing a significant energy gain over the non-magnetized one. Interestingly, a clear peak is observed in energy distribution for a magnetized target with $\sim$ 1.45 MeV at the rear surface. This indicates the excitation of mono-energetic electrons in the magnetized target. The neodymium target is 5 mm thick in the experiment and under such conditions, we cannot observe the highly energized electrons reaching the rear surface.
 
To study the behavior of electrons that have gained energy from the front surface during laser interaction. We randomly selected 2000 electrons of energy greater than 0.5 MeV (from the front surface up to $2 \mu m$ depth) and tracked their path for 64 fs. Figure \ref{fig:Current_trajectory_sim}(a) and \ref{fig:Current_trajectory_sim}(b) illustrate the tracking history of these particles for non-magnetic and magnetic targets respectively. Clearly, the magnetized target collimates the electron propagation along the magnetic field.  In contrast, the electron flow is random and divergent for non-magnetic targets. The electrons gyrating with the magnetic field are essentially tied to the field lines.     Figure \ref{fig:Current_trajectory_sim}(c) and \ref{fig:Current_trajectory_sim}(d) present net electron current density 64 femtoseconds laser irradiates the target. Certain notable differences between magnetic and non-magnetic targets are (a) localization of currents and (b) deeper penetration of electrons into the over-dense plasma for the former. This results in currents appearing on the target rear. 

As the electrons get ejected from the front and back surfaces of the plasma, they generate a sheath region. This has been shown by plotting the electric fields in figure \ref{fig:sheath_field} for both kinds of targets. Higher magnitude of the sheath field depicts that higher number of electrons ejected from the magnetic target. This sheath field prevents further ejection of electrons from the plasma target surface. One important observation (figure \ref{fig:sheath_field}(b)) is that the sheath field is comparatively higher in the bulk and rear surfaces of the magnetized target. 
While the number of electrons with higher energy increases with the magnetic field, the maximum energy cut-off remains typically the same with increasing magnetic field. This has been illustrated in Fig \ref{fig:cut_off_energy} which provides a comparison for two values of 
applied external B fields (viz., 2kT, and 4kT). For all these values, the laser-plasma interaction is limited only to the plasma surface as the laser is unable to penetrate the overdense plasma target (the unmagnetized dispersion relation is still applicable). At a high magnetic field of 20kT, the laser propagation is governed by the magnetized dispersion relation and is found to be in the pass band. We observe from the electron energy spectra plot of Fig. \ref{fig:energy_dist_time_evol} that there is a considerable increase in the electron energy in this case. 


\subsection{Multiple mono-energetic electron beam generation}

 Interestingly, the electron energy spectra show well-defined peaks (see Figure \ref{fig:Sim_Spectrum}) in simulations for the magnetic targets at certain values of the magnetic field. These peaks essentially illustrate the monoenergetic electron beam generation in the vicinity of certain energies. We have explored this aspect further by carrying out systematics at three different values of the magnetic field for different times as shown in Fig.\ref{fig:energy_dist_time_evol}$(a,b,c)$. The plot of the energy spectra for the magnetic field value of 2 kT shows the most prominent development of peaks. This particular feature is absent for non-magnetic targets and for magnetic targets at a very high field of 20kT. A closer examination reveals that the peaks occur at energy locations for which the ratio of relativistically modified electron Larmor radius and the laser wavelength $r_e/\lambda $ (or the ratio of laser frequency to the relativistically modified gyro-frequency $(\omega_l/\omega_{ce}) \gamma$ ) is close to integer values, as depicted in Figure \ref{fig:mono_energy}. Here $r_e$ is the relativistically modified gyro-radius of the electrons evaluated for the electron energy at which the peaks appear. It is also important to note that these peaks in the spectra appear at a much later time when the laser has already left. It thus indicates that the imprint of the laser characteristics is still present in the plasma in the form of some excited modes. 

 The absence of the peaks at higher magnetic fields can easily be understood. At high magnetic fields, $r_e/\lambda $ continues to remain fractional even at the highest energy of the electrons observed in the spectrum. We feel that the appearance of these peaks can be traced to the excitation of electrostatic mode in the plasma by the laser. The mode should have a frequency similar to that of the incident laser. Those particles, that have energy typically near the range where their relativistically modified gyro frequency is close to integral multiples of the frequency of electrostatic mode, would get excited by the wave and keep acquiring energy. We are working on quantifying this in a better fashion to develop a theoretical model of this phenomenon.
 
\section{Conclusions}
 We have demonstrated enhancement of collimation, flux and energy of relativistic electron pulses generated  by an intense, femtosecond laser pulse on the application of a  modest, readily available magnetic field of 0.1 tesla.  This is in contrast to the many previous simulations and a few experimental tests where 100s-1000s Tesla have been used for the same purpose. 2D PIC simulations reproduce our experimental data very well. In addition, they uncover a new feature namely, the generation of a quasi-monochromatic energy bunch at the high end of the energy spectrum. Even more interesting is the tunability and optimization of the energy location of this bunch by the applied magnetic field. We offer an  explanation for the energy location in terms of simple plasma dynamics. Our results  should be of interest to several areas like intense laser-driven electromagnetic radiation and particle sources, fast ignition of laser fusion and laboratory  astrophysics. \\

\noindent\begin{Large}
\textbf{Funding}.\end{Large} GRK acknowledges partial support from J.C. Bose Fellowship grant (JBR/2020/000039) from the Science and Engineering Board (SERB), Government of India. AD acknowledges support from  the SERB core grants CRG 2018/000624 and CRG/2022/002782 as well as a J C Bose Fellowship grant JCB/2017/000055. TD would like to thank IIT Delhi HPC facility for computational resources and the Council for Scientific and Industrial Research (Grant No-09/086/(1489)/2021-EMR-I) for funding the research. ADL acknowledges support from the Infosys-TIFR Leading Edge Research Grant (Cycle 2)\\

\noindent\begin{Large}
\textbf{Acknowledgement}\end{Large}
We acknowledge Virender Ranga from the TIFR Department of nuclear and atomic physics for the GEANT4 simulation of electron propagation through Neodymium (NdFeB) magnetic material.
\bibliography{references}
\bibliographystyle{ieeetr}

\end{document}